\documentclass[letterpaper]{JHEP3}
\usepackage{epsfig}
\usepackage{graphicx}
\usepackage{subfigure}
\usepackage{dcolumn}
\usepackage{bm}
\usepackage{float}

\newcommand{\beq}{\begin{equation}}
\newcommand{\eeq}{\end{equation}}
\newcommand{\bq}{\begin{equation}}
\newcommand{\eq}{\end{equation}}
\newcommand{\ba}{\begin{array}}
\newcommand{\ea}{\end{array}}
\newcommand{\beqa}{\begin{eqnarray}}
\newcommand{\eeqa}{\end{eqnarray}}

\def\O{{\cal O}}

\def\End{\end{document}}

\def\to{\rightarrow}

\def\dis{\displaystyle}
\def\f{\frac}
\def\ov{\overline}
\def\[{\left[}
\def\]{\right]}
\def\({\left(}
\def\){\right)}

\def\DL{\Delta {\mathcal L}}
\def\a{\alpha}
\def\b{\beta}
\def\GA{\Gamma}
\def\C{\mathcal C}
\def\U1EM{U(1)_{\rm em}}
\def\O{\mathcal O}

\def\[{\left[}
\def\]{\right]}
\def\dis{\displaystyle}

\def\cut{\Lambda}

\preprint{MADPH--09-1552}

\title{
\vspace*{0.75cm}
{\Large ${\mbox {\boldmath  ${\mu \tau}$}}$ Production at Hadron Colliders }
}
\author{Tao Han\footnote{than@hep.wisc.edu}, Ian Lewis\footnote{ilewis@wisc.edu}\\
{\it Department of Physics, University of Wisconsin, Madison, WI 53706, U.S.A.} }
\author{Marc Sher\footnote{mtsher@wm.edu} \\
{\it Particle Theory Group,~College of William and Mary, 
Williamsburg, Virginia 23187}}

\date{\today}

\abstract{
\noindent
Motivated by large $\nu_\mu-\nu_\tau$ flavor mixing, we consider  $\mu \tau$ production at hadron colliders 
via dimension-6 effective operators, which can be attributed to new physics in the flavor sector at a higher scale 
$\Lambda$. Current bounds on many of these operators from low energy  experiments are very weak or nonexistent, 
and they may lead to clean  $\mu^+\tau^-$ and $\mu^- \tau^+$ signals at  hadron colliders. 
At the Tevatron with 8 fb$^{-1}$, one can exceed current bounds for most operators, with most  2$\sigma$ 
sensitivities being in the $6-24$~TeV range.
We find that at the LHC with 1 fb$^{-1}$ (100 fb$^{-1}$) integrated luminosity, one can reach a
 $2\sigma$ sensitivity for $\Lambda\sim 3-10$~TeV ($\Lambda\sim 6-21$~TeV), 
 depending on the Lorentz structure of the operator.  For some operators, an improvement of several 
 orders of magnitude in sensitivity can be obtained with only a few tens of pb$^{-1}$ at the LHC.
 }


\begin{document}

\section{Introduction}

The most important discovery in particle physics in the past decade has only deepened the mystery of  ``flavor" of quarks and leptons.
The fact that the mixing angles in the leptonic sector are large \cite{atm,sol} stands in sharp contrast with the observed small mixing angles in the quark sector.   In particular, mixing between the second and third generation neutrinos appears to be maximal.  Of course, this large mixing could occur from diagonalizing the neutrino mass matrix, the charged lepton mass matrix, or both.   At present, the source of this large mixing is a mystery.  

In view of this, it is tempting to explore other interactions which change lepton flavor between the second and third generations.  Several years ago, two of us (TH, MS), along with Black and He (BHHS) \cite{black}, performed a  comprehensive analysis of constraints  on these interactions
based on low energy meson physics.   
BHHS chose an effective field theory approach, in which all dimension-6  operators of the form 
\begin{equation}
\label{eq:4F-def}
\( \bar\mu\, \Gamma\, \tau \)
( \bar{q}^\a \,\Gamma\, {q}^\b ) \,,
\end{equation}
were studied, where $\Gamma$ contains possible Dirac $\gamma$-matrices.    With six flavors of  quarks, there were 12 possible combinations of $q^a$ and $q^b$ (assuming Hermiticity), six diagonal and six off-diagonal, and four choices $S,P,V,A$ of the gamma matrices were considered.    All of these operators were considered, and most were bounded by considering $\tau$, $K$, $B$ and $t$ decays.  

In particular, BHHS considered operators of the form
\beq
\label{eq:taumu6}
\DL = 
\DL^{(6)}_{\tau\mu} 
=\dis\sum_{j,\alpha,\beta}
\f{~\C^j_{\alpha\beta}~}{\cut^2}
\left( \ov{\mu} ~\Gamma_j\, \tau \right) 
\left( \ov{q}^\alpha \,\Gamma_j\, q^\beta \right)
\,+\,{\rm H.c.}\,,
\eeq
where 
$\GA_j\in 
\(1,\,\gamma_5^{~},\,\gamma_\sigma^{~},\,\gamma_\sigma^{~}\gamma_5^{~}\)$ 
denotes relevant Dirac matrices, specifying scalar, pseudoscalar,
vector and axial vector couplings, respectively.    They did not consider tensor operators since the hadronic matrix 
elements were not known and the bounds were expected to be weak in any event.
They chose a value of 
\beq
\label{eq:C-def} 
\C^j_{\a\b}
            ~=~ 4\pi \,\O(1)   \,  ~~~~~({\rm default})\,,
\eeq
which corresponds to an underlying theory with a strong gauge coupling of $\alpha_S = \O(1)$.    Arguments can be made for multiplying or dividing this by $4\pi$, for naive dimensional analysis or for weakly coupled theories, respectively.   A discussion is found in BHHS; we simply choose the above definition of $\Lambda$ and other choices can be made by simple rescaling.

Besides the four fermion operators in Eq.~(\ref{eq:taumu6}), there may be other induced operators involving the SM gauge bosons, such as the electroweak transition operator
\beq
\label{eq:EWtrans}
\DL=\frac{\kappa v}{\Lambda^2}\bar{\mu}\sigma^{\mu\nu}\tau F_{\mu\nu},
\eeq
where $v$ is the vacuum expectation value of the Standard Model Higgs field and $F_{\mu\nu}$ is the electroweak field tensor.
However, when these operators are compared to the underlying new strong dynamics of the four fermion interaction in Eq.~(\ref{eq:taumu6}), it is found that they are suppressed by $\mathcal{O}(M_W/\Lambda)$, where $M_W$ is the mass of the electroweak gauge boson.  For new physics scales of order $1$ TeV or greater, this is at least an order of magnitude suppression.  Thus, we ignore these operators.

BHHS found that operators involving the three lightest quarks were strongly bounded, with bounds ranging from $3$ to $13$ TeV on the related value of $\Lambda$.  These bounds can be found in Appendix~\ref{app:bnds}.    Not surprisingly, operators involving the top quark were either unbounded or very weakly bounded, with only the $\ov{t}u$ operator for vector and axial vector couplings  being bounded by $\Lambda < 650$ GeV (the bound arises through a loop in $B\rightarrow \mu\tau$ decay).   Operators involving the $b$-quark and a light quark also have bounds on $\Lambda$ which were generally in the several TeV range.    However, there were some surprises.   The scalar and pseudoscalar operators involving $\ov{c}u$ and $\ov{c}c$ were completely unbounded, and the $\ov{b}b$ operator was essentially unbounded for all $S,P,V,A$ operators.   And, as noted above, {\bf none} of the tensor operators were considered at all, for all quark combinations.

In this note, we point out that the operators in Eq.~(\ref{eq:4F-def}) (without involving top 
quarks) will contribute to $\mu-\tau$ production at hadron colliders.    
Given that many of the possible operators, as noted above, are completely unbounded or weakly bounded from the current low energy data, 
study of $pp\rightarrow \mu\tau$ at the LHC or $p\ov{p}\rightarrow\mu\tau$ at the Tevatron will probe unexplored territory.    

There have been some previous discussions of $\mu-\tau$ production at hadron colliders.   Han and Marfatia \cite{danny} looked at the lepton-violating decay $h\rightarrow\mu\tau$ at hadron colliders, and a very detailed analysis of signals and backgrounds was carried out by 
Assamagan et al.~\cite{assamagan} afterwards.   
Other work looking at Higgs decays focused on mirror fermions \cite{diaz},  supersymmetric 
models \cite{rossi}, seesaw neutrino models \cite{arganda}, and Randall-Sundrum models \cite{azatov}.   In addition to Higgs decays, others have considered lepton-flavor violation in the decays of supersymmetric particles \cite{deppisch} and in horizontal gauge boson models \cite{bengtsson}.   These analyses, however, were done in the context of very specific models (often relying on the assumption that the $\mu$ and $\tau$ are emitted in the decay of a single particle).   Here, we will use a much more general effective field theory approach.

This paper is organized as follows.  In the next section, we discuss the cross sections for $\mu\tau$ production via 
the various operators.   A detailed analysis of the signal identification and background subtraction is in Section 3, 
and Section 4 contains some discussions and our conclusions.  
Appendix~\ref{app:bnds} reiterates the bounds from BHHS for comparison, and 
Appendix~\ref{app:unitarity} outlines the calculation of partial-wave unitarity bounds.

\section{$\mu\tau$ Production at Hadron Colliders}

Due to the absence of appreciable $\mu\tau$ production in the SM, their production can be estimated via
the effective operators in Eq.~(\ref{eq:4F-def}). 
On dimensional grounds, the cross section for $\bar{q}_iq_j\rightarrow\mu\tau$ grows with center of mass energy, i.e.,
\begin{eqnarray}
\sigma(\bar{q}_iq_j\rightarrow \mu\tau)\propto\frac{{s}}{\Lambda^4},
\end{eqnarray}
where $\sqrt{{s}}$ is the center of mass energy  for the partonic system.  This growth of cross section with 
energy  will eventually violate unitarity bounds.  
Expanding the scattering amplitudes in partial waves, 
we find the unitarity bounds to be (see Appendix~\ref{app:unitarity})
\begin{eqnarray}
s  \leq  \left\{  
\begin{array}{ll}
2\Lambda^2  & {\rm for\ scalar,\  pseudoscalar,\ and\ tensor;} \\
3\Lambda^2  & {\rm vector\ and\ axial~vector\  case.}
\end{array}
\right.
\label{eq:Unitarity}
\end{eqnarray}

The total cross sections for $\mu\tau$ production at the hadronic level after convoluting with the parton distribution 
functions (pdfs) are
\begin{eqnarray}
\sigma_{\rm Scalar}&=&\frac{\pi}{3}\frac{S}{\Lambda^4}\int^{\tau_{\rm max}}_{\tau_0} d\tau(q\otimes q)(\tau)\bigg{(}1-\frac{\tau_0}{\tau}\bigg{)}^2\tau\\
\sigma_{\rm Vector}&=&\frac{4\pi}{9}\frac{S}{\Lambda^4}\int^{\tau_{\rm max}}_{\tau_0} d\tau (q\otimes q)(\tau)\bigg{(}1-\frac{\tau_0}{\tau}\bigg{)}^2\bigg{(}1+\frac{\tau_0}{2\tau}\bigg{)}\tau\\
\sigma_{\rm Tensor}&=&\frac{8\pi}{9}\frac{S}{\Lambda^4}\int^{\tau_{\rm max}}_{\tau_0} d\tau (q\otimes q)(\tau)\bigg{(}1-\frac{\tau_0}{\tau}\bigg{)}^2\bigg{(}1+2\frac{\tau_0}{\tau}\bigg{)}\tau,
\end{eqnarray}
where $\tau=s/S$, $\tau_0=m^2_\tau/S$, $m_\tau$ is the tau mass, and $\sqrt{S}$ is the center of mass energy
in the lab frame.  
The pseudoscalar cross section is of the same form as the scalar cross section, and the axial vector cross section is of the same form as the vector cross section.  
 Our perturbative calculation will become invalid at the unitarity bound, hence there is a maximum on the $\tau$ integration.  It is given by
$\tau_{\rm max}=2\Lambda^2/S$ for the scalar, pseudoscalar, and tensor cases, and $\tau_{\rm max}=3\Lambda^2/S$ for the vector and axial-vector cases.
Also, $q(x)$ is the quark distribution function with flavor sum suppressed, and $\otimes$ denotes the convolution defined as
\begin{eqnarray}
(g_1\otimes g_2)(y)=\int^1_0dx_1\int^1_0dx_2\ g_1(x_1)g_2(x_2)\delta(x_1x_2-y) .
\end{eqnarray}
The CTEQ6L parton distribution function set is used for all of the results \cite{Stump:2003yu}.

Results for the cross sections for the scalar, pseudoscalar, vector, axial vector, and tensor structures at the Tevatron, LHC at 10 TeV and 14 TeV are given in Table~\ref{tab:LHCscal}.  The cross section for the pseudoscalar (axial vector) current is the same as for the scalar (vector) current.  For all cases, $\Lambda$ is set equal to 2 TeV and the unitarity bounds are taken into consideration.  At this rather high scale, the production rates are dominated by the valence quark contributions. The cross sections at the LHC are larger than those at the Tevatron by roughly an order of magnitude, reaching about 100 pb.

For some cases the bounds from BHHS are greater than $2$ TeV, hence the cross section needs to be scaled to determine a realistic cross section at hadron colliders.  The partonic cross sections scale at $\Lambda^{-4}$, but at the hadronic level a complication arises since the unitarity bounds introduce a dependence on the new physics scale in the integration over pdfs.  If the unitarity bounds are ignored ($\tau_{\rm max}=1$), one finds that with $\Lambda=2$~TeV neglecting the unitarity bounds has at most a 10\% effect on the cross sections at the LHC for both 10 TeV and 14 TeV and no effect at the Tevatron since the unitarity bounds are greater than the lab frame energy.  Hence, if $\Lambda$ is increased from 2 TeV, at the LHC it is a good approximation to assume the cross section scales as $\Lambda^{-4}$ and at the Tevatron the cross section scales exactly as $\Lambda^{-4}$.  For example, the lower bound on $\Lambda$ for the vector $u\bar{u}$ coupling from BHHS is 12 TeV, so the maximum cross section at the 14 TeV LHC from this operator would be approximately $160\times(2/12)^4$~pb = $120$ fb.  On the other hand, there is no bound whatsoever for the vector $u\bar{c}$ coupling, and thus a cross section limit of $110$~pb would yield a new limit of $2$~TeV on the scale of this operator.  This would constitute an improvement of many orders of magnitude. 

\begin{table}[tb]
\caption{Cross sections for all the scalar, pseudoscalar, vector, axial vector, and tensor structures at the Tevatron at 2 TeV, the LHC at 10 TeV, and the LHC at 14 TeV.  The pseudoscalar (axial vector) cross section is the same as the scalar (vector) cross section.  All cross sections were evaluated with the new physics scale $\Lambda = 2$~TeV and the unitarity bounds are taken into consideration.} 

\begin{center}
\begin{tabular}{|c|p{0.4in}|p{0.43in}|p{0.4in}|p{0.4in}|p{0.43in}|p{0.4in}|p{0.4in}|p{0.43in}|p{0.4in}|}  \hline
               &\multicolumn{3}{|c|}{Tevatron 2 TeV $(p\bar p)$}  &\multicolumn{3}{|c|}{LHC 10 TeV $(pp)$}
               &\multicolumn{3}{|c|}{LHC 14 TeV $(pp)$}\\ \hline
$\sigma $ (pb) & 1,$\gamma_5$  & $\gamma_\mu$,$\gamma_\mu\gamma_5$  & $\sigma_{\mu\nu}$            & 1,$\gamma_5$  & $\gamma_\mu$,$\gamma_\mu\gamma_5$  & $\sigma_{\mu\nu}$       & 1,$\gamma_5$  & $\gamma_\mu$,$\gamma_\mu\gamma_5$  & $\sigma_{\mu\nu}$\\ \hline
 $u\bar u$     & 8.4    &  11     & 22                & 63        & 85    & 170         & 120             & 160    & 310    \\ \hline
 $d\bar d$     & 2.5    &  3.3    & 6.7               & 38        & 51    & 100         & 72              & 98     & 190     \\ \hline
 $s\bar s$     & 0.18   &  0.24   & 0.49              & 5.5       & 7.4   & 15          & 11              & 15     & 30        \\ \hline
 $d\bar s$     & 1.3    &  1.7    & 3.4               & 34        & 45    & 91          & 66              & 89     & 180    \\ \hline
 $d\bar b$     & 0.50   &  0.67   & 1.3               & 17        & 22    & 45          & 34              & 46     & 90       \\ \hline
 $s\bar b$     & 0.13   &  0.17   & 0.34              & 5.0       & 6.7   & 13          & 11              & 14     & 28        \\ \hline
 $u\bar c$     & 1.5    &  2.0    & 3.9               & 41        & 55    & 110         & 80              & 110    & 210        \\ \hline
 $c\bar c$     & 0.070  &  0.094  & 0.19              & 2.6       &3.5    & 7.0         & 5.5             & 7.3    & 15        \\ \hline
 $b\bar b$     & 0.021  &  0.028  & 0.056             & 1.1       & 1.5   & 2.9         & 2.4             & 3.2    & 6.4        \\ \hline
\end{tabular}
\end{center}
\label{tab:LHCscal}
\end{table}

\section{Signal Identification and Backgrounds}
Upon production at hadron colliders, $\tau$'s will promptly decay and are detected via their decay products.  
About 35\% of the time the $\tau$ decays to two neutrinos and an electron or muon, the other 65\% 
of the time the $\tau$ decays to a few hadrons plus a neutrino.  
We will consider the $\tau$ decay to an electron as well as hadronic decays in this work. 
The decay to a muon will result in a $\mu^+\mu^-$ final state that has a large Drell-Yan background.
We will study the signal reach at the Tevatron and at the
14 TeV LHC.

\subsection{$\tau$ Decay to Electrons }
\subsubsection{Signal Reconstruction}
The $\tau$ decays to an electron plus two neutrinos about 18\% of the time.  We thus search for a final state of an electron and muon

\begin{equation}
e + \mu .
\label{eq:emu}
\end{equation}

The electromagnetic calorimeter resolution is simulated by smearing the electron energies 
according to a Gaussian distribution with a resolution parameterized by 
\begin{equation}
\frac{\sigma(E)}{E}=\frac{a}{\sqrt{E/{\rm GeV}}}\oplus b ,
\label{eq:enres}
\end{equation}
where the constants are $a=10\%$ and $b=0\%$ at the Tevatron \cite{Carena:2000yx}, $a=5\%$ and $b=0.55\%$ at the LHC \cite{Ball:2007zza},
and $\oplus$ indicates addition in  quadrature.
For simplicity, we have used the same form of smearing for the muons.

\begin{figure}[tb]
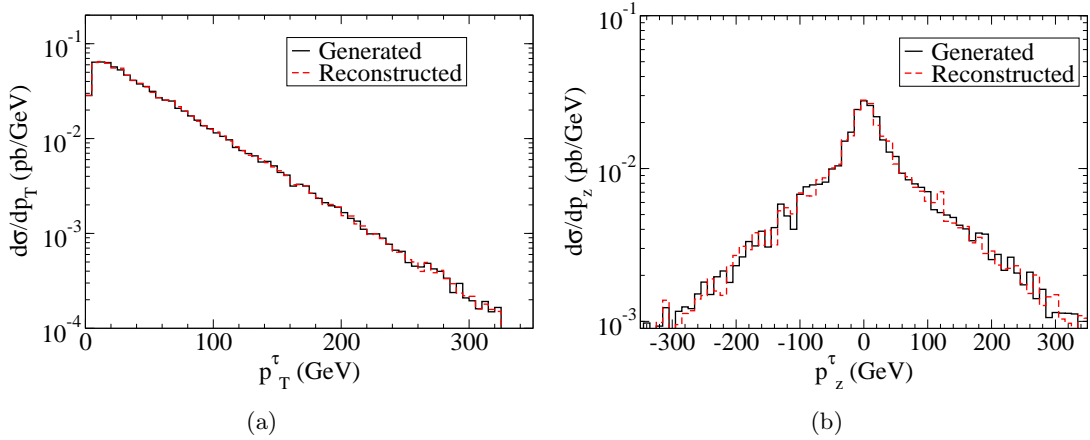

\centering
\subfigure[]{
	\label{fig:taurecT}
	\includegraphics[width=0.45\textwidth]{pttauucbar.eps}
	}
\subfigure[]{
	\includegraphics[width=0.45\textwidth]{pztauucbar.eps}
	\label{fig:taurecz}
}

\caption{Distributions of the theoretically generated (solid line) and kinematically reconstructed (dashed line) $\tau$ momentum at the Tevatron at 2 TeV with a $u\bar{c}$ initial state, scalar coupling, and new physics scale of $1$~TeV.
Fig. (a) is the $\tau$ transverse momentum distribution, and Fig. (b) is the $\tau$ longitidunal momentum distribution.}
\label{fig:taurec}
\end{figure}

The decay of the $\tau$ leaves us with some missing energy and we need to consider how to effectively
reconstruct the $\tau$ momentum.  
For our process all the missing transverse momentum is coming from the $\tau$, hence 
\begin{eqnarray}
\mathbf{p}^\tau_T=\mathbf{p}^{e}_T+\mathbf{p}^{\rm miss}_T .
\label{eq:recon1} 
\end{eqnarray}
At hadron colliders, we have no information on the longitudinal component of the missing momentum on an event-by-event
basis.  
However, the $\tau$ will be highly boosted and its decay products will be collimated.  Hence, the missing momentum should be aligned with the electron momentum and the ratio $p^{e}_z/p^{\rm miss}_z$ should be the same as the ratio of the magnitudes of the transverse momenta, $p^{e}_T/p^{\rm miss}_T$.  Therefore, the longitudinal component of the $\tau$ can be reconstructed as \cite{danny}
\begin{eqnarray}
p^\tau_z=p^{e}_z\bigg{(}1+\frac{p^{\rm miss}_T}{p^{e}_T}\bigg{)} .
\label{eq:recon2}
\end{eqnarray}
Once the three-momentum is reconstructed, we can solve for the $\tau$ energy, $E^2_\tau=\mathbf{p}^2_\tau+m^2_\tau$.  Figure \ref{fig:taurec} illustrates the effectiveness of this method at the Tevatron.  
Figure \ref{fig:taurecT} (Figure \ref{fig:taurecz}) shows the transverse momentum (longitudinal momentum) distribution for the theoretically generated (solid) and kinematically reconstructed (dashed) $\tau$ momenta.  As can be seen, the $\tau$ momentum is reconstructed effectively.

We first apply some basic cuts on the transverse momentum and the pseudo rapidity
to simulate the detector  acceptance and triggering, as well as to isolate the signal from the background, 
\begin{eqnarray}
&p^\mu_T>20~{\rm GeV},~~~~~~~&|\eta^\mu|<2.5, 
\nonumber\\
&p^{e}_T>20~{\rm GeV},~~~~~~~&|\eta^{e}|<2.5 .
\label{eq:cuts1}
\end{eqnarray}
Since the signal does not contain any jets, we also require a jet veto such that there are no jets with 
$p_T>50$~GeV and $|\eta|<2.5$.

There are several distinctive kinematic features of our signal.   The decay products of the $\tau$ will be highly collimated, and the electron transverse momentum will be traveling in the same direction as the missing transverse momentum.  Also, in the transverse plane  the muon and tau should be back to back.  Since the electron will mostly be in the direction of the $\tau$, it will also be nearly back to back with the muon. Finally, the $\tau$ and $\mu$ have equal transverse momenta; hence, the decay products of the $\tau$ have less transverse momentum than the $\mu$.  We can measure this discrepancy using the momentum imbalance
\begin{equation}
\Delta p_T=p^\mu_T-p^{e}_T.
\end{equation}
For the signal, this observable should be positive.  Based on the kinematics of our signal, we apply the further cuts \cite{assamagan}
\begin{eqnarray}
\delta\phi(p^\mu_T,p^{e}_T)&>&2.75~{\rm rad},~~~~~ \delta\phi(p^{\rm miss}_T,p^{e}_T)~<~0.6~{\rm rad},
\label{eq:cuts2}\\
\Delta p_T&>&0.\nonumber
\end{eqnarray}

\subsubsection{Backgrounds and their Suppression}
The leading backgrounds are $W^+W^-$ pair production, $Z^0/\gamma^\star\rightarrow \tau^+\tau^-$, and $t\bar{t}$ pair production \cite{assamagan}. 
The total rates for these backgrounds at the Tevatron and the LHC are given in Table~\ref{tab:tevabginit}
with consecutive cuts. 
We consider both of the final states with $\mu^+$ and $\mu^-$.

\begin{table}[tb]
\caption{Leading backgrounds to the $\tau$'s electronic  decay  before and after consecutive kinematic and invariant mass cuts for (a) the Tevatron at 2 TeV and (b) the LHC at 14 TeV.}
\begin{center}
\begin{tabular}{|c|c|c|c|c|}  \hline
~~Backgrounds (pb)  & No Cuts & Cuts Eq.~(\ref{eq:cuts1}) & + Eq.~(\ref{eq:cuts2}) & + Eq.~(\ref{eq:invcut}) \\
\hline
\multicolumn{5}{|c|}{(a) Tevatron 2 TeV}    \\ \hline 
$W^+W^-\rightarrow \mu^{\pm}\nu_\mu\tau^{\mp}\nu_\tau$                    &0.032         &0.0046 &0.0012  &$2.6\times10^{-4}$ \\
\hline
$W^+W^-\rightarrow \mu^{\pm}\nu_\mu e^{\mp}\nu_e$                    &   0.18           &0.13   &0.0060 &$9.8\times10^{-4}$   \\
\hline
$ Z^0/\gamma^\star\rightarrow \tau^+\tau^-\rightarrow \mu^{\pm}\nu_\mu\tau^{\mp}$&610     &0.21   &0.091      & $1.4\times10^{-4}$       \\
\hline
$ t\bar{t}\rightarrow \mu^{\pm}\nu_\mu b\tau^{\mp}\nu_\tau\bar{b}$         & 0.020        &$6.5\times10^{-4}$    &$7.4\times10^{-5}$ &$4.4\times10^{-5}$\\
\hline
$ t\bar{t}\rightarrow \mu^{\pm}\nu_\mu b e^{\mp}\nu_e\bar{b}$         &   0.11           &0.0099     &$7.3\times10^{-4}$  &$2.7\times10^{-4}$\\
\hline

\multicolumn{5}{|c|}{(b) LHC 14 TeV}    \\ \hline
$W^+W^-\rightarrow \mu^{\pm}\nu_\mu\tau^{\mp}\nu_\tau$                    & 0.34     &0.030      & 0.0088 & 0.0031         \\
\hline
$W^+W^-\rightarrow \mu^{\pm}\nu_\mu e^{\mp}\nu_e$                    &1.9       &0.99      & 0.051       & 0.014  \\
\hline
$Z^0/\gamma^\star\rightarrow \tau^+\tau^-\rightarrow \mu^{\pm}\nu_\mu\tau^{\mp}$&2300     &1.1        &0.49      & 0.0014       \\
\hline
$t\bar{t}\rightarrow \mu^{\pm}\nu_\mu b\tau^{\mp}\nu_\tau\bar{b}$         &1.9    &0.070     &0.010 &0.0077\\
\hline
$ t\bar{t}\rightarrow \mu^{\pm}\nu_\mu b e^{\mp}\nu_e\bar{b}$         & 11    & 1.5    &0.10  &0.050\\
\hline
\end{tabular}
\end{center}
\label{tab:tevabginit}
\end{table}

The partonic cross section of our signal increases with energy while the cross sections of the backgrounds 
will decrease with energy.  
Hence, the invariant mass distribution of our signal does not fall off as quickly as the backgrounds.  

\begin{figure}[tb]
\centering
\subfigure[]{
	\label{fig:tevaimindiv}
	\includegraphics[width=0.36\textwidth,angle=-90]{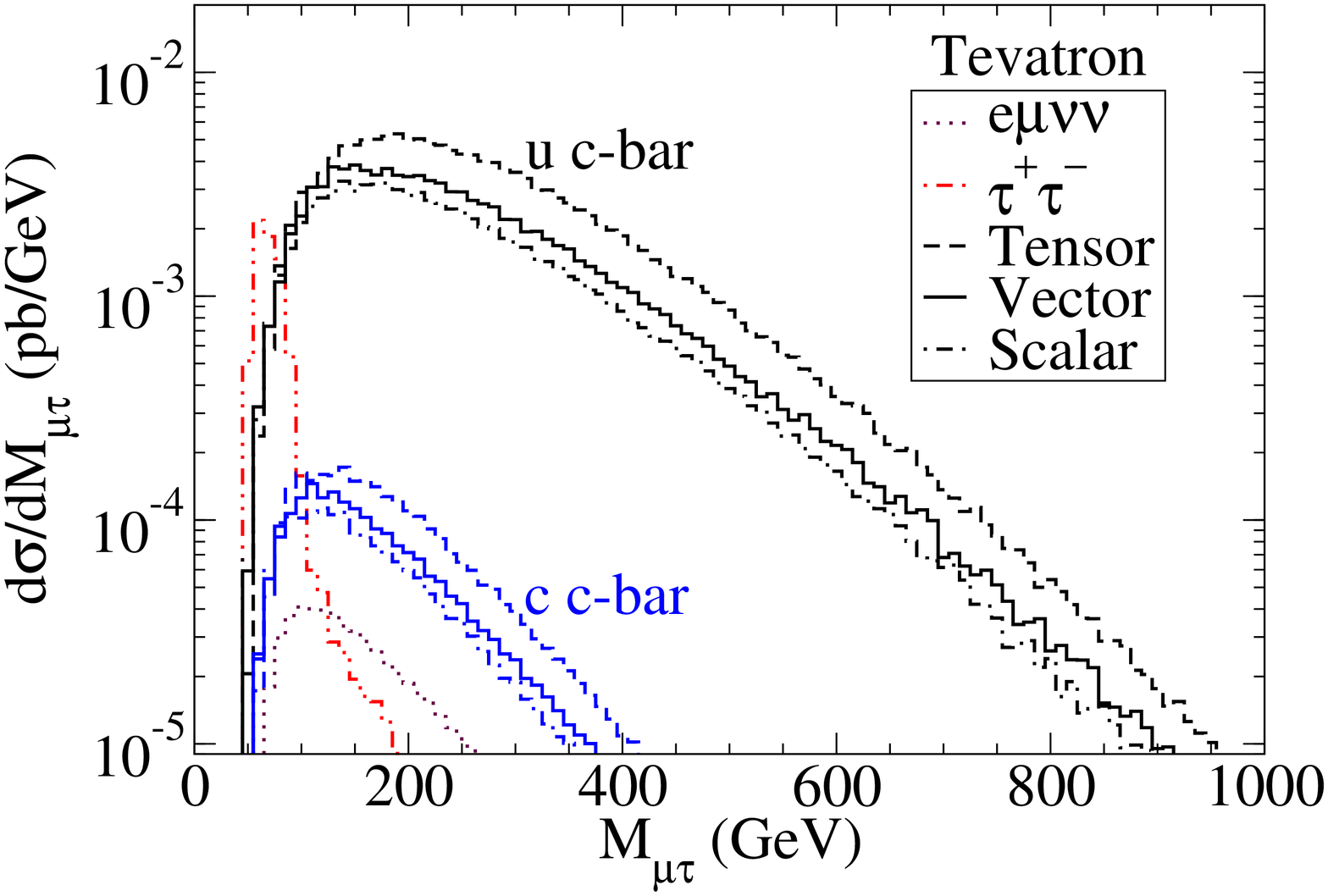}
	}
\subfigure[]{
	\includegraphics[width=0.36\textwidth,angle=-90]{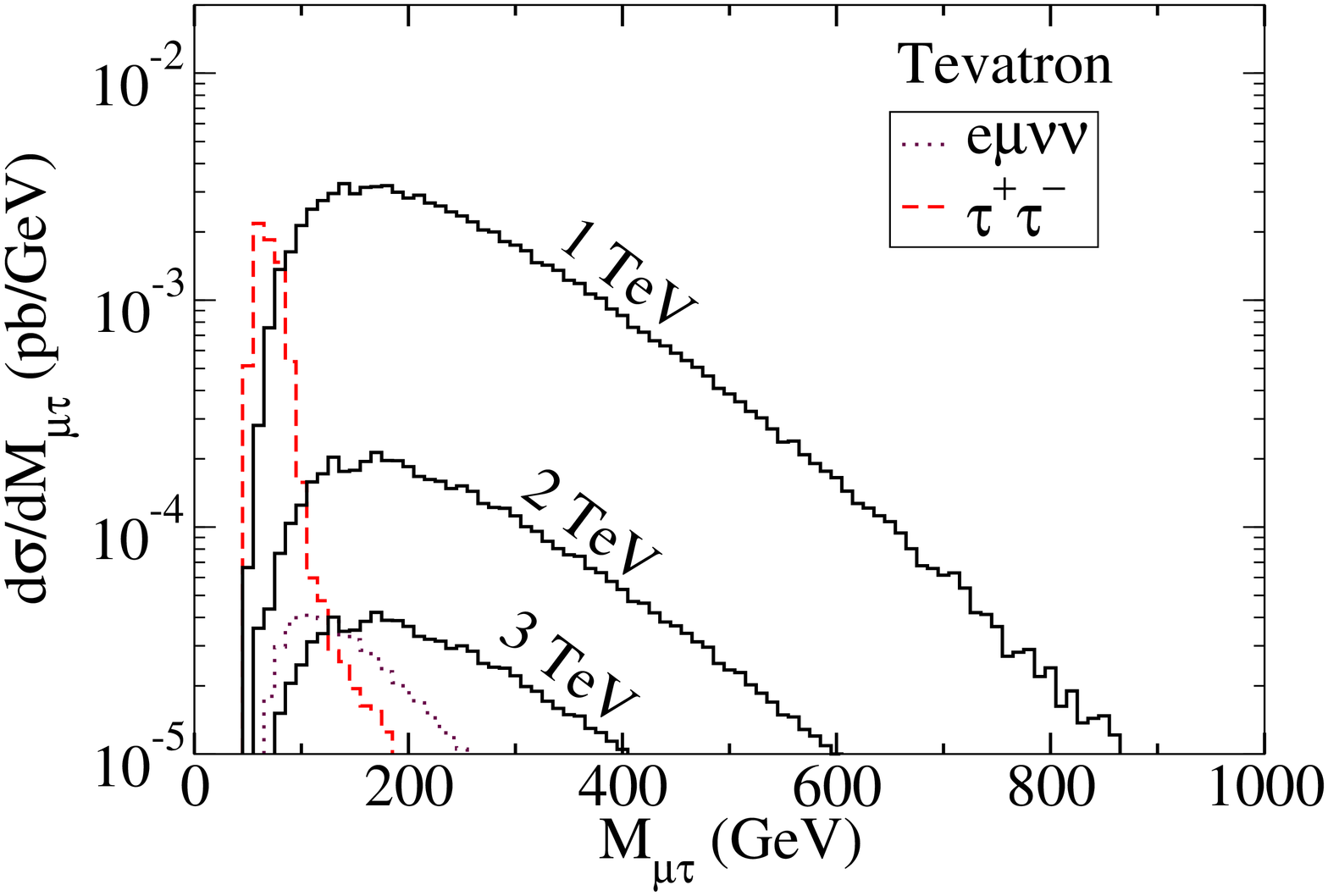}
	\label{fig:tevaimvary}
}

\caption{The invariant mass distributions of the reconstructed $\tau-\mu$ system at the Tevatron at 2 TeV.  Fig. (a) shows the distributions of the leading backgrounds (dotted and dot-dot-dash) and of our signal for the $u\bar c$ and $c\bar c$ initial states with coupling of various Lorentz structures and a new physics scale of $1$~TeV.  Fig. (b) shows the distributions of the leading backgrounds (dotted and dashed) and of our signal (solid) for the $u\bar{c}$ initial state with scalar coupling and various new physics scales.  The cuts in Eqs. (3.5) and (3.7) have been applied.}
\label{fig:tevainvmass}
\end{figure}

Figure \ref{fig:tevaimindiv} shows the invariant mass distributions of backgrounds
and our signal at the Tevatron with initial states $c\bar c$ and $u\bar c$ with various couplings and a new physics scale of $1$~TeV after applying the cuts in Eqs.~(\ref{eq:cuts1}) and (\ref{eq:cuts2}).  
The cross section for the pseudoscalar (axial-vector) couplings are the same as those for the scalar (vector) couplings.
The decline in the signal rates is due to a suppression of the pdfs at large $x$.  Although the signal rates steeply decline with invariant mass the background falls off faster.  The $u\bar c$ signal is still clearly above background due to a valence quark in the initial state, but the $c\bar c$ signal distribution is much closer to the background distribution due to the steep fall with invariant mass and a lack of an initial state valence quark.  
Figure \ref{fig:tevaimvary} shows the invariant mass distributions of backgrounds and our signal at the Tevatron with initial state $u\bar c$ and scalar coupling for various new physics scales.  The 3 TeV new physics scale invariant mass distribution is approaching the background distribution. A higher cutoff on the invariant mass will be needed to separate the weak 
signal from the backgrounds. Based on Fig.~\ref{fig:tevainvmass}, we propose a selection cut on 
\begin{equation}
M_{\mu\tau} >250~{\rm GeV.} \label{eq:invcut}
\end{equation}
Table~\ref{tab:tevabginit} shows the effects of the invariant mass cut on the backgrounds in the last column.

\begin{figure}[tb]
\centering
\subfigure[]{
	\label{fig:imindiv}
	\includegraphics[width=0.36\textwidth,angle=-90]{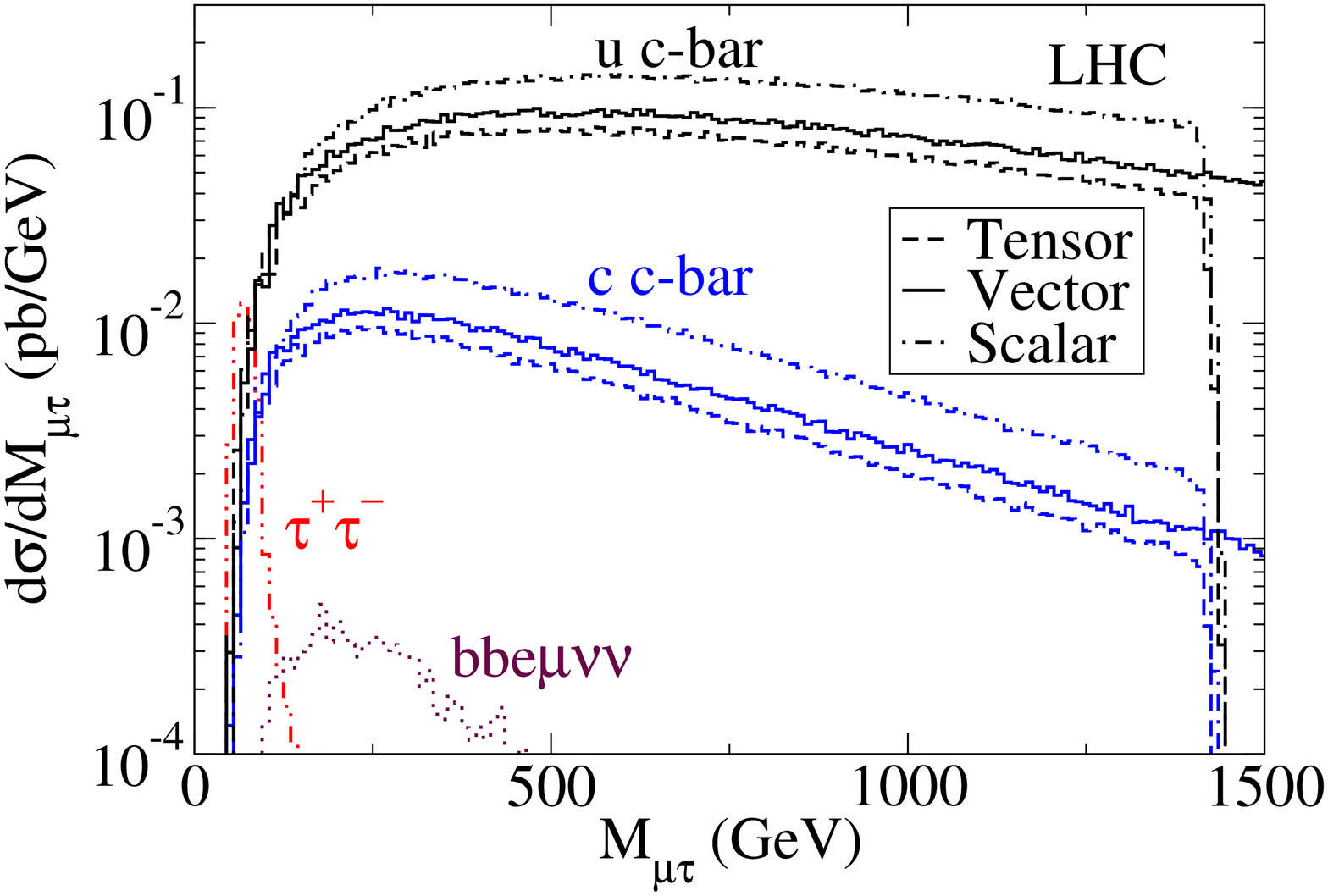}
	}
\subfigure[]{
	\includegraphics[width=0.36\textwidth,angle=-90]{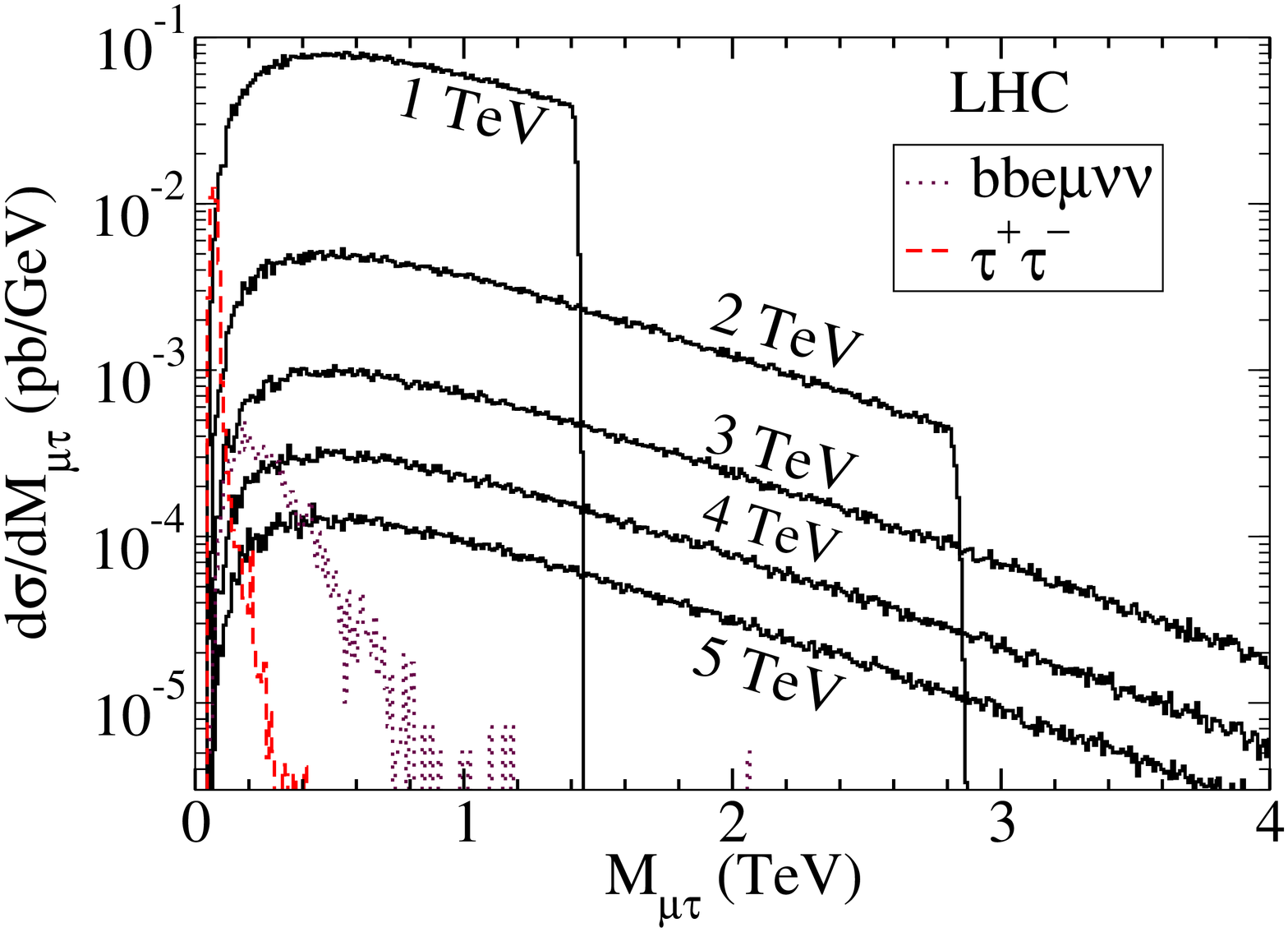}
	\label{fig:imvary}
}

\caption{The invariant mass distributions of the reconstructed $\tau-\mu$ system at the LHC at 14 TeV.  Fig. (a) shows the distributions of the leading backgrounds (dotted and dot-dot-dash) and of our signal for the $u\bar c$ and $c\bar c$ initial states with coupling of various Lorentz structures and a new physics scale of $1$~TeV.  Fig. (b) shows the distributions of the leading backgrounds (dotted and dashed) and of our signal (solid) for the $u\bar{c}$ initial state with scalar coupling and various new physics scales. The cut offs in the distributions at high invariant mass are due to the unitarity bounds.  The cuts in Eqs. (3.5) and (3.7) have been applied.}
\label{fig:invmass}
\end{figure}

Similar analyses can be carried out for the LHC. 
Figure \ref{fig:imindiv} shows the invariant mass distribution for our signal with the $u\bar c$ and $c\bar c$ initial states and various Lorentz structures, as well as the backgrounds after the cuts in Eqs.~(\ref{eq:cuts1}) and (\ref{eq:cuts2}).
The new physics scale was set to $1$~TeV and the unitarity bound is imposed.  
Figure \ref{fig:imvary} shows the invariant mass distribution of the $u\bar{c}$ initial state with various new physics scales.  The cutoff on the invariant mass corresponds to the unitarity bound, the scale at which the perturbative calculation becomes untrustworthy.  In the lack of the knowledge for the new physics to show up at the scale $\Lambda$, we simply impose a
sharp cutoff at the unitarity bound.   
As compared with the Tevatron, the LHC signal rates fall off much less quickly with invariant mass since the Tevatron's lower energy leads to a suppression from the pdfs at large $x$. As can be seen, as the new physics scale increases the cross section decreases and the background becomes more problematic at lower invariant mass.  Also, as the new physics scale increases the unitarity bound becomes less strict.  Hence, although the backgrounds at the LHC are considerably larger than at the Tevatron, for large new physics scales the LHC has an enhancement in the signal cross section from the large invariant mass region.

\subsection{$\tau$  Decay to Hadrons}

Although with significantly larger backgrounds, the signal from $\tau$ hadronic decays can be very distinctive as well.
We limit the hadronic $\tau$ decays to 1-prong decays to pions, i.e., $\tau^{\pm}\rightarrow\pi^{\pm}\nu_\tau$, $\tau^{\pm}\rightarrow\pi^{\pm}\pi^0\nu_\tau$, and $\tau^{\pm}\rightarrow \pi^{\pm}2\pi^0\nu_\tau$.  The $\tau$'s have 1-prong decays to these final states about 50\% of the time. We thus search for a final state of a $\tau$ jet and a muon
\begin{equation}
j_{\tau} + \mu.
\label{eq:jetmu}
\end{equation}

To simulate detector resolution effects, the energy is smeared according to Eq.~(\ref{eq:enres}) with $a=80\%$ and $b=0\%$ for the jet at the Tevatron \cite{Carena:2000yx} and $a=100\%$ and $b=5\%$ at the LHC \cite{Ball:2007zza}.   As in the electronic decay, the $\tau$ is highly boosted and its decay products are collimated.    Hence, all the missing energy in the event should be aligned with the $\tau$.   The signal is then reconstructed as described in Eqs.~(\ref{eq:recon1}) and (\ref{eq:recon2}) with the electron momentum replaced by the momentum of the $\tau$-jet.

The hadronic decay of the $\tau$ also has the backgrounds $W^+W^-$ pair production, $Z^0/\gamma^\star\rightarrow \tau^+\tau^-$, and $t\bar{t}$ pair production plus an additional background of $W+$jet, where the jet is misidentified as a $\tau$-jet. At the Tevatron, we assume a $\tau$-jet tagging efficiency of $67\%$ and that a light jet is mistagged as a $\tau$-jet $1.1\%$ of the time \cite{Svoisky:2009zz} and at the LHC we assume a $\tau$-jet tagging efficiency of $40\%$ and a light jet misidentification rate of $1\%$ \cite{Ball:2007zza}.
Even with a low rate of misidentification, the $W+$jet background is large.  To suppress this background, we note that for hadronic decays most of the $\tau$ transverse momentum will be carried by the jet.  Hence the $\tau$-jet should be traveling in the same direction as the reconstructed $\tau$ momentum. Motivated by this observation, we apply the same cuts as Eqs.~(\ref{eq:cuts1}), (\ref{eq:cuts2}), and (\ref{eq:invcut}) with the electron momentum replaced by the $\tau$-jet momentum and the additional cuts
\begin{eqnarray}
\frac{p^{\tau-jet}_T}{p^\tau_T}&>&0.6~~~~~~~~~~~~~\Delta R(p^{\tau-jet}_T,p^\tau_T)~<~0.2~{\rm rad}.~~~\label{eq:hadcuts}
\end{eqnarray}

\subsection{Sensitivity Reach at the Tevatron}

One can determine the sensitivity of the Tevatron to the new physics scale with 8 fb$^{-1}$ of data.  
Table~\ref{tab:tevasignalsenshad} shows the sensitivity of the Tevatron for (a) electronic 
and (b) hadronic $\tau$ decays. The tables list the maximum new physics scale sensitivity at 
2$\sigma$ and 5$\sigma$ level at the Tevatron.
The reaches for scalar (vector) and pseudoscalar (axial-vector) are the same at the Tevatron, although the previous bounds from BHHS for the scalar (vector) and pseudoscalar (axial-vector) couplings may not be the same. The bounds from BHHS can be found in Appendix~\ref{app:bnds}.  If only one of the bounds for scalar (vector) or pseudoscalar (axial-vector) coupling from BHHS is greater than the Tevatron reach one star is placed next to the new physics scale, if both bounds are greater than the Tevatron reach two stars are placed next to the new physics scale.  Due to the larger backgrounds from $W+$jet, the Tevatron is much less sensitive to the $\tau$ hadronic decays than the $\tau$ electronic decays.

  There were no bounds from BHHS for the tensor couplings, so the Tevatron will be able to exlude some of the parameter space.  Since the tensor cross sections are generally at least twice as large as the scalar cross sections, the Tevatron is more sensitive to the tensor couplings than it is to scalar couplings.  Also, in general, the Tevatron is more sensitive to processes with initial state valence quarks than those without initial state quarks.
With 8 fb$^{-1}$ of data most of the bounds can be increased, some quite stringently.

Somewhat similar leptonic final states have been searched for 
in a model-independent way
at the Tevatron \cite{Abbott:2000fb}, although these included 
substantial missing energy and possible jets.  We encourage the 
Tevatron experimenters to carry out the analyses as suggested
in this article.

\begin{table}[tb]
\caption{Maximum new physics scales the Tevatron is sensitive to with 8 fb$^{-1}$ of data at the 2$\sigma$ and 5$\sigma$ levels.  The sensitivities are presented for both (a) electronic and (b) hadronic $\tau$ decays with various initial states.
One star indicates that the Tevatron reach is less than only one of the scalar (vector) or pseudoscalar (axial-vector) bounds from BHHS, and two stars indicates that the Tevatron reach is less than both bounds from BHHS.  BHHS does not contain bounds on the tensor coupling.}
\begin{center}
\begin{tabular}{|c|p{0.4in}|p{0.44in}|p{0.4in}|p{0.4in}|p{0.44in}|p{0.4in}|}  \hline
\multicolumn{7}{|c|}{(a) $\tau \to e$ }    \\ \hline 
$\Lambda_{\rm{NP}}$ (TeV)&\multicolumn{3}{|c|}{2$\sigma$ sensitivity}   &  \multicolumn{3}{|c|}{5$\sigma$ discovery} \\ \hline 
Coupling      & 1,$\gamma_5$  & $\gamma_\mu$,$\gamma_\mu\gamma_5$  & $\sigma_{\mu\nu}$        & 1,$\gamma_5$  & $\gamma_\mu$,$\gamma_\mu\gamma_5$  & $\sigma_{\mu\nu}$      \\
\hline 
$u\bar u$     &20       &21         & 24        &14       &15       &17    \\
\hline
$d\bar d$     &17       &18         & 21        &12       &13       &15   \\
\hline
$s\bar s$     &9.9      &10         & 12        &7.2*     &7.7**    &8.7      \\
\hline
$d\bar s$     &15       &16         & 18        &10       &11*      &13   \\
\hline 
$d\bar b$     &13       &14         & 16        &9.8      &10*      &11   \\
\hline 
$s\bar b$     &9.5      &10         & 11        &6.9      &7.3      &8.3   \\
\hline 
$u\bar c$     &17       &18         & 20        &12       &13       &14   \\
\hline
$c\bar c$     &7.9      &8.3        & 9.5       &5.7      &6.0     &6.9   \\
\hline
$b\bar b$     &6.4      &6.8        & 7.7       &4.6      &4.9      &5.6  \\
\hline

\multicolumn{7}{|c|}{(b) $\tau \to h^{\pm}$ }    \\ \hline 
$\Lambda_{\rm{NP}}$ (TeV)  &\multicolumn{3}{|c|}{2$\sigma$ sensitivity}   &  \multicolumn{3}{|c|}{5$\sigma$ discovery}\\ \hline 
Coupling      & 1,$\gamma_5$  & $\gamma_\mu$,$\gamma_\mu\gamma_5$  & $\sigma_{\mu\nu}$       & 1,$\gamma_5$  & $\gamma_\mu$,$\gamma_\mu\gamma_5$  & $\sigma_{\mu\nu}$   \\
\hline 
$u\bar u$     &8.6**    &9.2**   &10      &6.5**   &6.9**  &8.1\\
\hline
$d\bar d$     &5.7**    &6.1**   &7.1     &4.3**   &4.6**   &5.4\\
\hline
$s\bar s$     &1.8*     &1.9**   &2.3     &1.4**   &1.4**   &1.7   \\
\hline
$d\bar s$     &3.7      &4.0*    &4.6     &2.8*    &3.0**   &3.5  \\
\hline 
$d\bar b$     &2.7*     &2.9*    &3.4     &2.0**   &2.2     &2.5  \\
\hline 
$s\bar b$     &1.5**    &1.6**   &1.9     &1.1**   &1.2**   &1.4   \\
\hline 
$u\bar c$     &3.9      &4.1     &4.8     &2.9     &3.1     &3.6  \\
\hline
$c\bar c$     &1.1      &1.2     &1.4     &0.89    &0.95**  &1.1 \\
\hline 
$b\bar b$     &0.91     &0.97    &1.1     &0.68    &0.73    &0.86\\
\hline
\end{tabular}
\end{center}
\label{tab:tevasignalsenshad}
\end{table}

\subsection{Sensitivity Reach at the LHC}

\begin{table}[h!]
\caption{Maximum new physics scales the LHC is sensitive to  at 14 TeV with 100 fb$^{-1}$ of data at the 2$\sigma$ and 5$\sigma$ levels.  The sensitivities are presented for both (a) electronic and (b) hadronic $\tau$ decays with various initial states. 
One star indicates that the LHC reach is less than only one of the scalar (vector) or pseudoscalar (axial-vector) bounds from BHHS, and two stars indicates that the LHC reach is less than both bounds from BHHS.  BHHS does not contain bounds on the tensor coupling.}
\begin{center}
\begin{tabular}{|c|p{0.4in}|p{0.44in}|p{0.4in}|p{0.4in}|p{0.44in}|p{0.4in}|}  \hline
\multicolumn{7}{|c|}{(a) $\tau \to e$ }    \\ \hline 
$\Lambda_{\rm{NP}}$ (TeV)  &\multicolumn{3}{|c|}{2$\sigma$ sensitivity}   &  \multicolumn{3}{|c|}{5$\sigma$ discovery} \\ \hline 
Coupling      & 1,$\gamma_5$  & $\gamma_\mu$,$\gamma_\mu\gamma_5$  & $\sigma_{\mu\nu}$       & 1,$\gamma_5$  & $\gamma_\mu$,$\gamma_\mu\gamma_5$  & $\sigma_{\mu\nu}$   \\
\hline 
$u\bar u$     &18      &19     &21     &14     &15     &17\\
\hline
$d\bar d$     &16      &17     &19     &12     &13     &15\\
\hline
$s\bar s$     &9.0*    &9.6*   &11     &7.1*   &7.6**  &8.6   \\
\hline
$d\bar s$     &13      &14     &16     &10     &11*    &13  \\
\hline 
$d\bar b$     &12      &13     &14     &9.7    &10     &11  \\
\hline 
$s\bar b$     &8.7     &9.2    &10     &6.8    &7.3    &8.2   \\
\hline 
$u\bar c$     &15      &16     &18     &12     &13     &14  \\
\hline
$c\bar c$     &7.2     &7.6    &8.6     &5.7    &6.0    &6.8 \\
\hline 
$b\bar b$     &5.8     &6.2    &7.0     &4.6    &4.9    &5.5\\
\hline
\multicolumn{7}{|c|}{(b) $\tau \to h^{\pm}$ }    \\ \hline 
$\Lambda_{\rm{NP}}$ (TeV)  &\multicolumn{3}{|c|}{2$\sigma$ sensitivity}   &  \multicolumn{3}{|c|}{5$\sigma$ discovery} \\ \hline 
$u\bar u$     &15      &16     &18     &12     &13     &14\\
\hline
$d\bar d$     &13      &14     &16     &10*    &11*    &13\\
\hline
$s\bar s$     &7.9*    &8.4**  &9.7    &6.2*   &6.7**  &7.7   \\
\hline
$d\bar s$     &11      &12*    &14     &9.3    &9.9*   &11  \\
\hline 
$d\bar b$     &10      &11     &13     &8.4*   &8.9    &10  \\
\hline 
$s\bar b$     &7.6     &8.1    &9.3    &6.0    &6.4    &7.4   \\
\hline 
$u\bar c$     &13      &14     &16     &10     &11     &12  \\
\hline
$c\bar c$     &6.3     &6.7    &7.8    &5.0    &5.3    &6.2 \\
\hline 
$b\bar b$     &5.1     &5.5    &6.3    &4.1    &4.3    &5.0\\
\hline
\end{tabular}
\end{center}
\label{tab:LHCsignalsensemu}
\end{table}
The LHC is also sensitive to flavor changing operators.  For the signal and background analysis, we used the same kinematical cuts as we used at the Tevatron, see Eqs.~(\ref{eq:cuts1}), (\ref{eq:cuts2}), and (\ref{eq:invcut}).  Table~\ref{tab:LHCsignalsensemu} shows the sensitivity of the LHC to all possible initial states and the couplings under consideration with 100 fb$^{-1}$ of data. The table contains the maximum new physics scales the LHC is sensitive to at the
2$\sigma$ and 5$\sigma$ levels.  As with the Tevatron, the LHC reach for scalar (vector) couplings is the same as that for pseudoscalar (axial-vector) couplings, although the bounds from BHHS may be different.  If only one of the bounds for scalar (vector) or pseudoscalar (axial-vector) coupling from BHHS is greater than the LHC reach one star is placed next to the new physics scale, if both bounds are greater than the LHC reach two stars are placed next to the new physics scale.  Despite the larger backrounds for the hadronic $\tau$ decays, at the LHC the reaches for the hadronic and electronic $\tau$ decays are much more similar than at the Tevatron since the LHC cross section receives an enhancement from the large invariant mass region.  For electronic (hadronic) $\tau$ decays the LHC with 100 fb$^{-1}$ of data is less (more) sensitive than the Tevatron with 8 fb$^{-1}$ of data.

\begin{figure}[ht]
\centering
\subfigure[]{
	\label{fig:ucbsens}
	\includegraphics[width=0.36\textwidth,angle=-90]{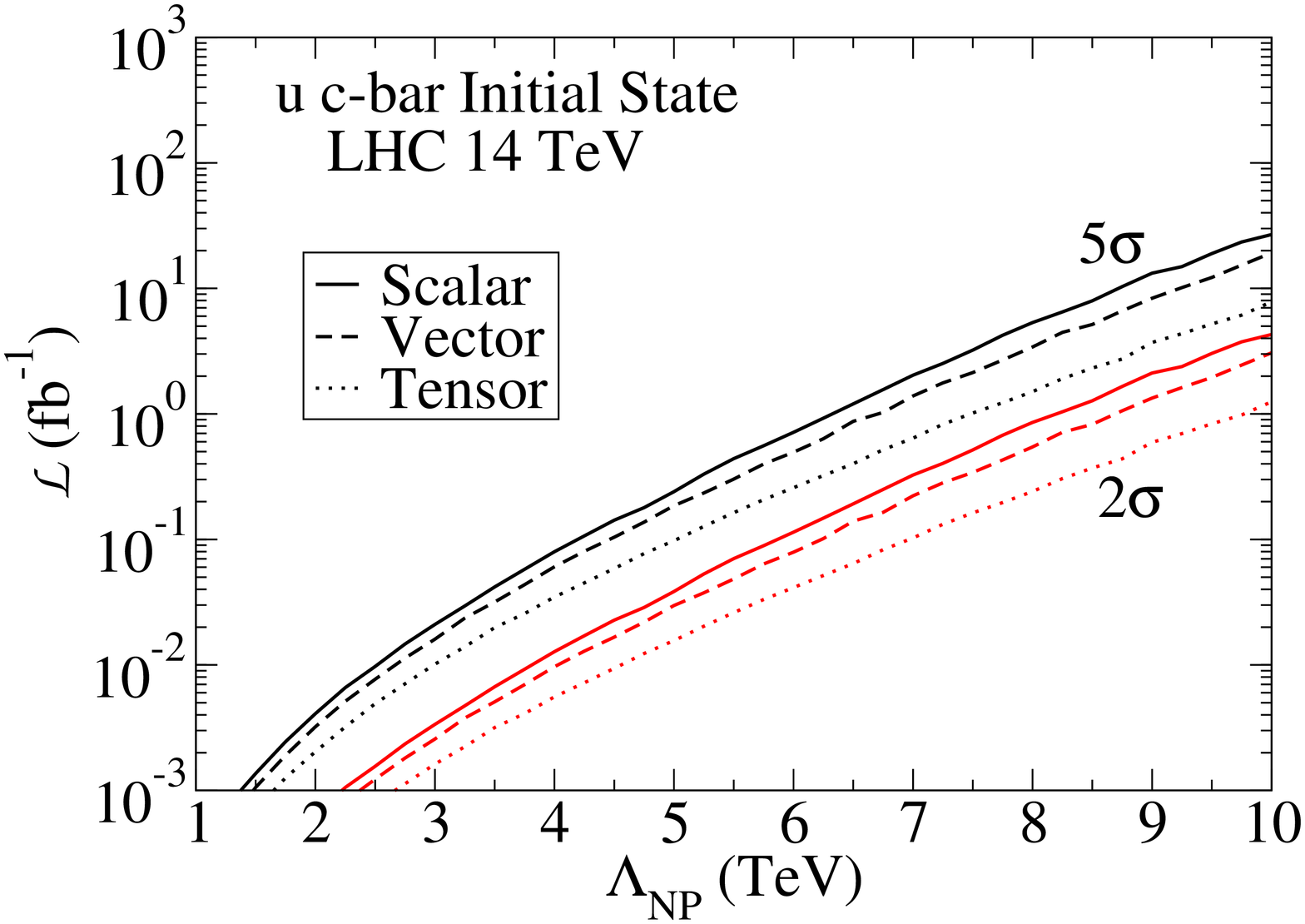}
	}
\subfigure[]{
	\includegraphics[width=0.36\textwidth,angle=-90]{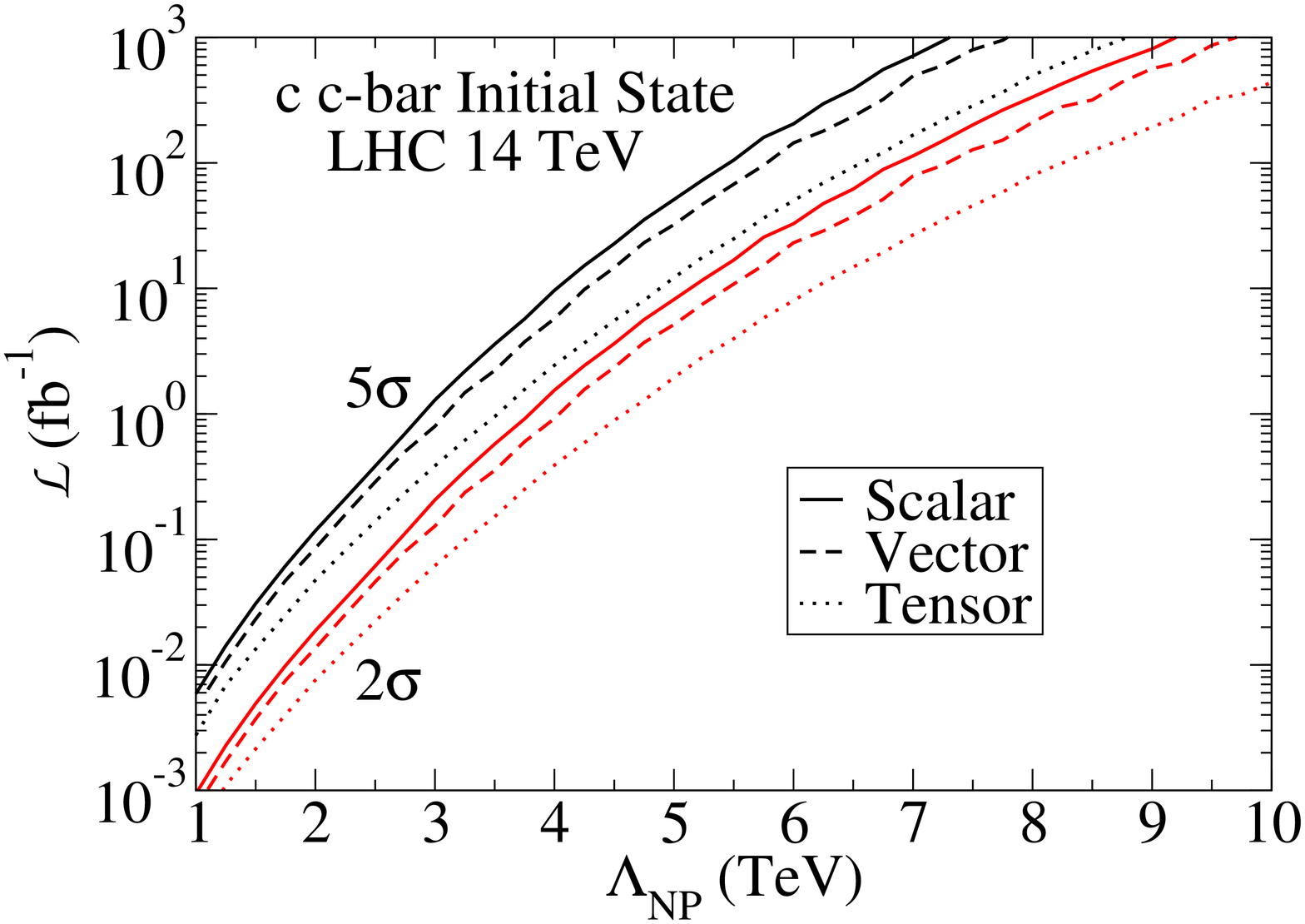}
	\label{fig:ccbsens}
}
\subfigure[]{
        \includegraphics[width=0.36\textwidth,angle=-90]{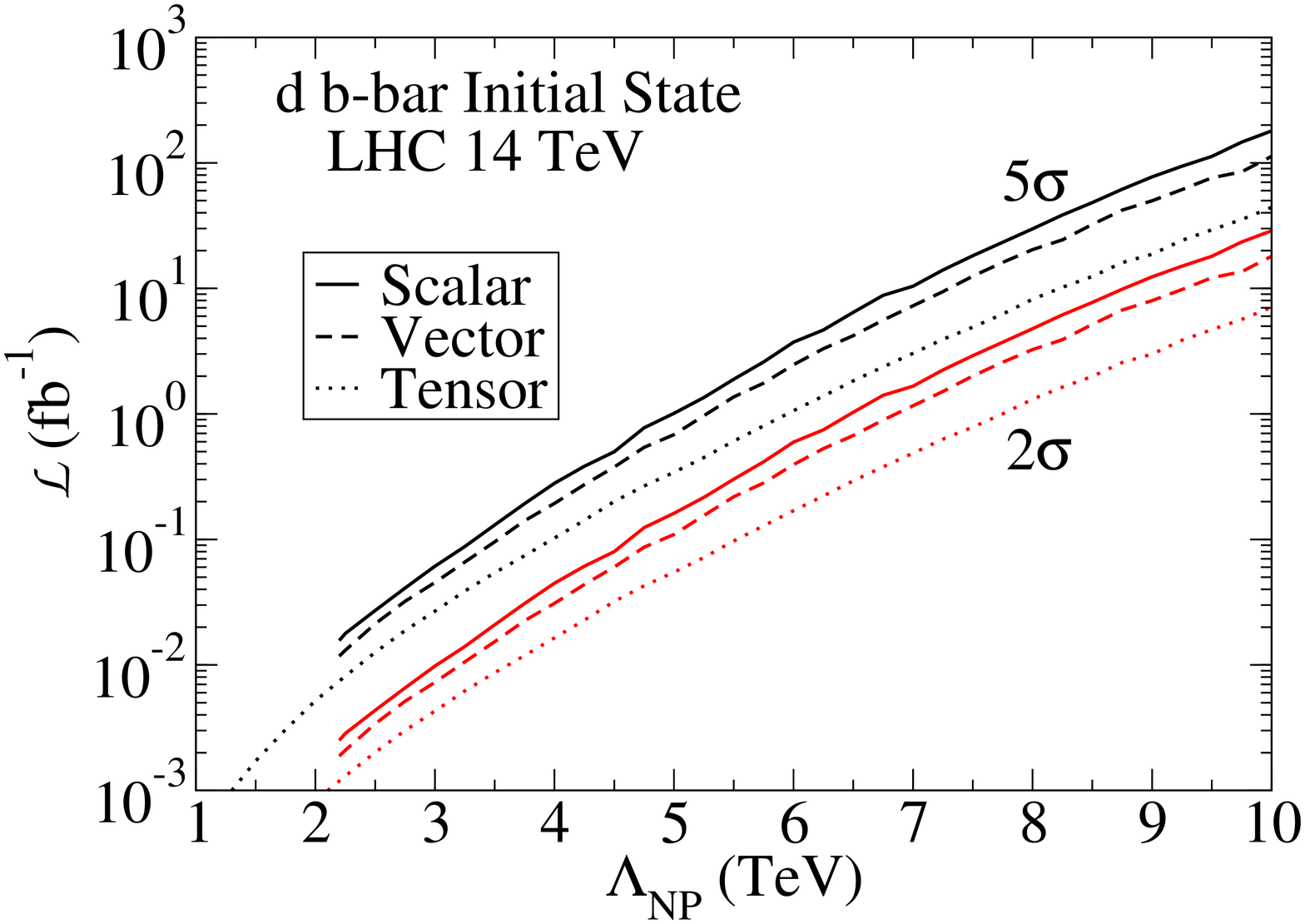}
        \label{fig:dbbsens}
}
\subfigure[]{
        \includegraphics[width=0.36\textwidth,angle=-90]{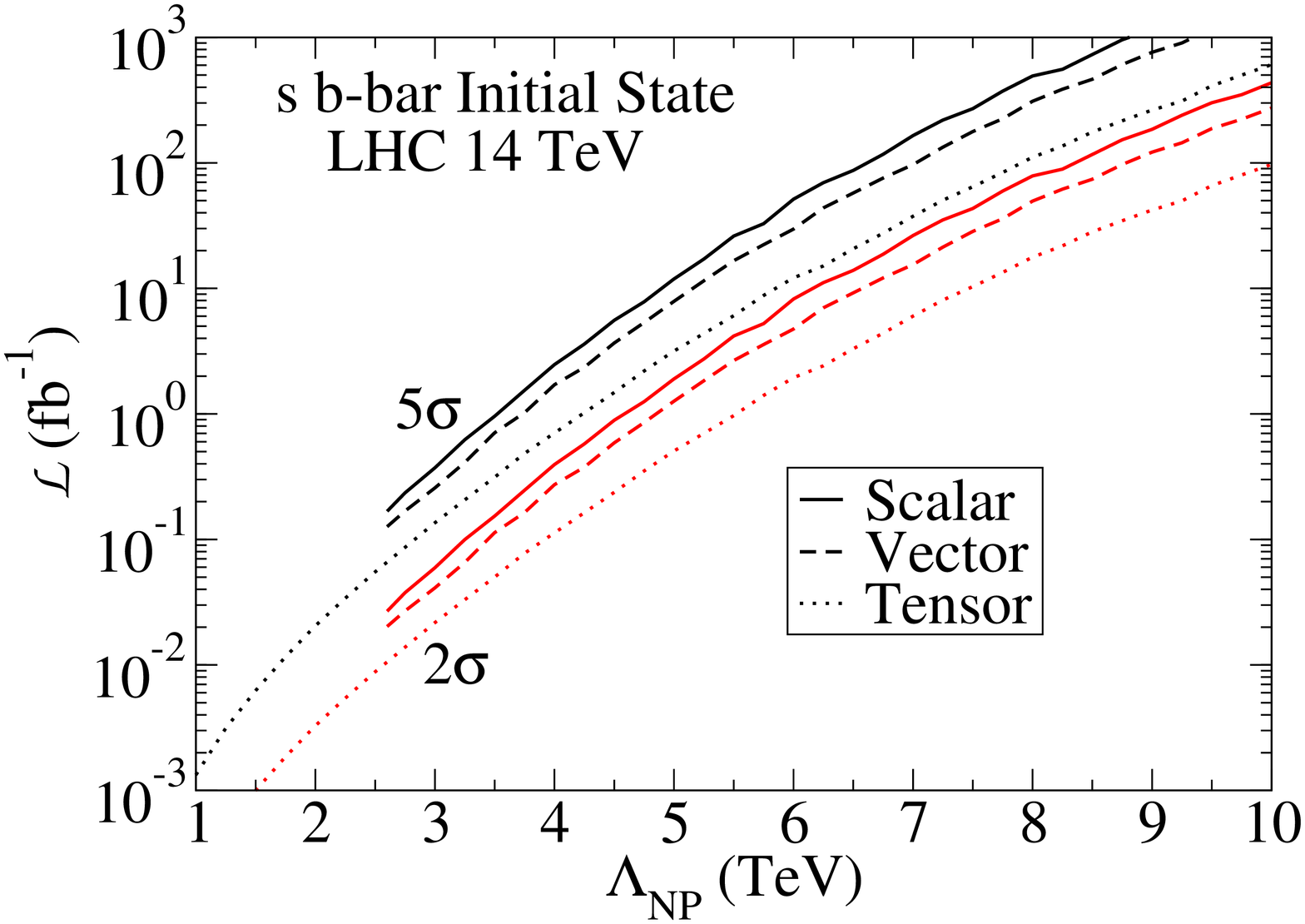}
        \label{fig:sbbsens}
}

\caption{The luminosity at the $14$ TeV LHC needed for $2\sigma$ and $5\sigma$ 
observation as a function of the new physics scales with couplings of various Lorentz structures and electronic $\tau$ decay. The sensitivity for the $u\bar{c}$ initial state is shown in 
(a), for the $c\bar{c}$ initial state in (b), for the $d\bar{b}$ initial state in (c), and for the $s\bar{b}$ initial state in (d).  The lower bounds on the new physics scale were taken from BHHS. }
\label{fig:LHCsens}
\end{figure}

Figure \ref{fig:LHCsens} shows the integrated luminosities needed for 2$\sigma$ and 5$\sigma$ observation at the LHC with various initial states and $\tau$ decay to electrons as a function of the new physics scale.  For some initial states and Lorentz structures BHHS had a bound on the new physics scale larger than $1$~TeV.  In those cases the distribution does not begin until the BHHS bound on the new physics scale.  The sensitivity for the pseudoscalar (axial-vector) is the same as the scalar (vector) state, although the bounds from BHHS are different. 
Note that extraordinary improvement in the bounds could be found (or a discovery made) with relatively low integrated luminosity.   Consider, for example, the $u\bar c$ initial state.   There is currently no bound at all; in principle, $\Lambda$ could be tens of GeV.   The figure shows that a total integrated luminosity of an inverse picobarn would give a $5\sigma$ sensitivity for a $\Lambda$ of $1$~TeV.   An integrated luminosity of an inverse femtobarn would give substantial improvements for all of the operators shown in Fig.~\ref{fig:LHCsens}.

 \section{Discussions and Conclusions}

In a previous article, motivated by discovery of large 
$\nu_\mu - \nu_\tau$  mixing in charged current interactions,  
bounds on the analogous mixing in neutral current interactions were explored.
A general formalism for dimension-6  fermionic effective operators involving
$\tau-\mu$ mixing with typical Lorentz structure
$(\ov{\mu} \,\Gamma \tau)(\ov{q}^\a \Gamma {q}^\b )$ was presented, and the low-energy constraints 
on the new physics scale associated with each operator were derived, 
mostly from experimental bounds 
on rare decays of $\tau$, hadrons or heavy quarks.   Tensor operators were not considered, and some of the operators, such as $\ov{c}u \ov{\mu}\tau$, were completely unbounded. 

In this article, we consider  $\mu \tau$ production at hadron colliders 
via these operators. 
Tables \ref{tab:tevasignalsenshad} and \ref{tab:LHCsignalsensemu}
list the new physics scales that are accessible at the Tevatron  and the LHC, respectively.  
Due to much smaller backgrounds, both the LHC and Tevatron are more sensitive to electronic $\tau$ decays than 
hadronic $\tau$ decays. For hadronic $\tau$ decays, the LHC receives an enhancement from the large invariant mass region and is more sensitive than the Tevatron.  Since the backgrounds to electronic $\tau$ decays at the Tevatron are much smaller than those at the LHC, the Tevatron is more sensitive than the LHC to electronic $\tau$ decays.    
We found that 
at the Tevatron with 8 fb$^{-1}$, one can exceed current bounds for most operators, with most  2$\sigma$ 
sensitivities being in the $6-24$~TeV range.
We find that at the LHC with 1 fb$^{-1}$ (100 fb$^{-1}$) integrated luminosity, one can reach a
 $2\sigma$ sensitivity for $\Lambda\sim 3-10$~TeV ($\Lambda\sim 6-21$~TeV), 
 depending on the Lorentz structure of the operator.

\subsection*{Acknowledgments}
We would like to thank Vernon Barger and Xerxes Tata for discussions. MS would 
like to thank
the Wisconsin Phenomenology Institute, in particular Linda Dolan, 
for hospitality during his visit.
The work  of TH and IL was supported by the US DOE under contract No.~DE-FG02-95ER40896,
and that of MS was supported in part by the National Science Foundation PHY-0755262.

\appendix

\section{New Physics Bounds}
\label{app:bnds}
The bounds from BHHS in units of TeV are presented in Table \ref{tab:BHHSbnds}. The *’s indicate there
are no bounds on the new physics scale. Also, there are no bounds from BHHS for the
tensor coupling.
\begin{table}
\begin{center}
\begin{tabular}{|c|p{0.8in}|p{0.8in}|p{0.8in}|p{0.8in}|}  \hline
Coupling type & 1     & $\gamma_5$& $\gamma_\mu$     & $\gamma_\mu\gamma_5$ \\
\hline
$u\bar u$     & 2.6       & 12          &12          &11 \\
\hline
$d\bar d$     & 2.6       &12           &12          &11  \\
\hline
$s\bar s$     &1.5        &9.9          & 14         &9.5     \\
\hline
$d\bar s$     &2.3        &3.7          & 13         &3.6    \\
\hline 
$d\bar b$     &2.2        &9.3          & 2.2        &8.2    \\
\hline 
$s\bar b$     &2.6        &2.8          & 2.6        &2.5     \\
\hline 
$u\bar c$     &*          &*            & 0.55       &0.55    \\
\hline
$c\bar c$     &*          &*           & 1.1        &1.1   \\
\hline
$b\bar b$     &*          &*           & 0.18       &*  \\
\hline
\end{tabular}
\end{center}
\caption{Bounds on the new physics scales from BHHS in units of TeV for various operators and
the scalar, pseudoscalar, vector, and axial-vector couplings. The *’s indicate there were no bounds.}
\label{tab:BHHSbnds}
\end{table}

\section{Partial Wave Unitarity Bounds}
\label{app:unitarity}
Since the cross section from our higher-dimensional operators increases as $s$, it is necessary to determine the unitarity bound for $q\bar{q}\rightarrow \mu\tau$.   The partial wave expansion for $a+b\rightarrow 1+2$ can be written as
$${\cal M}(s,t)=16\pi\sum_{J=M}^\infty (2J+1)a_J(s) d^J_{\mu\mu'}(\cos\theta)$$
where
$$a_J(s) = {1\over 32\pi}\int_{-1}^1 {\cal M}(s,t) d^J_{\mu\mu'}(\cos\theta) d\cos\theta,$$
$\mu=s_a-s_b$, $\mu'=s_1-s_2$ and $J\leq {\rm max}(|\mu|,|\mu'|)$.   The condition for unitarity is $|\Re(a_J)| \leq 1/2$.

It is straightforward to calculate the coefficients for the S,V,T operators.   For example, for the scalar operator $${\cal M} = {4\pi\over\Lambda^2}\bar{v}_{\lambda_1}(p_1)u_{\lambda_2}(p_2)\bar{u}_{\lambda_3}(p_3)v_{\lambda_4}(p_4)$$
one can just plug in the explicit expressions:
$$
u_\lambda(p)\equiv\left({\sqrt{E-\lambda |p|}\chi_\lambda(\hat{p}) \atop
\sqrt{E+\lambda |p|}\chi_\lambda(\hat{p})} \right) $$

$$v_\lambda(p)\equiv\left({-\sqrt{E+\lambda |p|}\chi_{-\lambda}(\hat{p}) \atop
\sqrt{E-\lambda |p|}\chi_{-\lambda}(\hat{p})} \right)
$$
where $\chi_+(\hat{z}) = \left({1\atop 0}\right), \chi_-(\hat{z})=\left({0\atop 1}\right)$.   In the massless limit, this simply gives $a_0 = s/(4\Lambda^2)$ and so the unitarity bound gives $s\leq 2\Lambda^2$.   For the vector case, $a_0 = 0$ and $a_1 = s/(6\Lambda^2)$ giving the unitarity bound $s\leq 3\Lambda^2$.  The tensor case gets contributions from both $a_0$ and $a_1$, and the stronger bound then applies.

\renewcommand{\baselinestretch}{1.4}

\end{document}